\documentclass[12pt]{article}
\usepackage{hyperref}
\usepackage{amsmath,amssymb}
\usepackage{graphicx}
\usepackage{color}
\newcommand{\Mat}[1]{{{\boldsymbol{#1}}}}
\newcommand{\abs}[1]{\left\vert#1\right\vert}
\definecolor{myblue}{rgb}{0.14,0.11,0.49}
\definecolor{myred}{rgb}{0.74,0.22,0.15}
\definecolor{mygreen}{rgb}{0.05,0.52,0.42}
\newcommand{\Couleur}[1]{\textcolor{myblue}{#1}}

\def\be{\begin{equation}}
\def\ee{\end{equation}}
\def\bea{\begin{eqnarray}}
\def\eea{\end{eqnarray}}
\def\bc{\begin{center}}
\def\ec{\end{center}}
\def\bi{\begin{itemize}}
\def\ei{\end{itemize}}
\def\dd{\operatorname{d}}
\def\iC{\operatorname{i}}
\def\noi{\noindent}
\begin{document}
\title{Interaction energy between a charged medium and its electromagnetic field as a dark matter candidate}
\author{
Mayeul Arminjon\\
\small\it Univ. Grenoble Alpes, CNRS, Grenoble INP
, 3SR, F-38000 Grenoble, France
}
\date{}
\maketitle
\begin{abstract}
In the scalar theory of gravitation with a preferred reference frame, a consistent formulation of electrodynamics in the presence of gravitation needs to introduce an additional energy tensor: the interaction energy tensor. This energy is gravitationally active and might contribute to the dark matter, because it has an exotic character and it is not localized inside matter. In order to check if that energy might form representative dark halos, one has to model the interstellar radiation field in a galaxy as a complete electromagnetic field obeying the Maxwell equations. A model has been built for this purpose, based on assuming axial symmetry and on recent results about axisymmetric Maxwell fields. Its predictions for the variation of the spectral energy density inside our Galaxy are relatively close to those of a recent radiation transfer model, except on the symmetry axis of the Galaxy, where the present model predicts extremely high values of the energy density. 
\end{abstract}

\section{Introduction}

Our initial motivation for the present work was independent of the problem of dark matter. It was to develop a consistent electrodynamics in an alternative theory of gravity: ``the scalar ether theory", or SET. This is a preferred-frame theory based on a scalar field only, \cite{A35, A59} that reduces to special relativity (SR) when the gravitational field vanishes. In general relativity (GR), the modification of the equations of electrodynamics in the presence of a gravitational field consists simply in rewriting the equations that are valid in SR, by using the ``comma goes to semicolon" rule:\Couleur{$\quad _{,\,\nu} \ \rightarrow \ _{;\,\nu}$},\ i.e.: partial derivatives are replaced by covariant derivatives based on the metric connection. (See Ref.~\cite{Stephani-Relativity} for an interesting discussion.) In particular, the dynamical equation for the energy(-momentum-stress) tensor \Couleur{$\Mat{T}$} that is valid in SR is: \  \Couleur{$T^{\lambda \nu }_{\ \, ,\nu}=0$}. Using the rule mentioned above, that equation is modified to:\   \Couleur{$ T^{\lambda \nu }_{\ \, ;\nu}=0$}, which is indeed the dynamical equation in GR and in many of its extensions or modifications. However, in the general situation, the latter equation is not equivalent to the dynamical equation of SET, \cite{A35} hence the foregoing rule cannot be used in SET.\\

Therefore, in that alternative theory, a different and less obvious path has to be taken for the purpose of adaptating classical electrodynamics in the presence of a gravitational field. It turns out that this leads to introduce an exotic form of energy, and that this new form is a possible candidate for dark matter. In this conference paper, we quickly follow that path. We then summarize the Maxwell model of the interstellar radiation field, that we built to prepare the test of this candidate. 

\section{Necessity of an interaction tensor in SET}

In SET, we assume classically that the electromagnetic field tensor \Couleur{$\Mat{F}$} derives from a 4-potential \Couleur{$A_\mu$}:
\be\label{Def F}
\Couleur{F_{\mu \nu } :=  A_{\nu ,\mu } - A_{\mu, \nu } = A_{\nu;\mu } - A_{\mu ; \nu }}.
\ee 
This is (locally) equivalent to assuming that (i) \Couleur{$\Mat{F}$} is antisymmetric (\Couleur{$F_{\mu \nu } = - F_{\nu \mu \, })$} and (ii) the first group of the Maxwell equations is satisfied:
\be\label{Maxwell 1}
\Couleur{F_{\lambda \mu \, ,\nu } + F_{\mu \nu ,\lambda } + F_{\nu \lambda ,\mu } \equiv F_{\lambda \mu \, ;\nu } + F_{\mu \nu ;\lambda } + F_{\nu \lambda ;\mu } = 0.}
\ee  
(The first equality in (\ref{Maxwell 1}) is indeed an identity due to the antisymmetry of the field tensor and to the symmetry of the metric connection.) Therefore, in SET, the first group of the Maxwell equations is left unchanged. In a first version of electrodynamics in the presence of a gravitational field in SET, the second group of the Maxwell equations was got by applying the dynamical equation of SET to a charged medium in the presence of the Lorentz force, assuming that the following holds for the energy tensors, as is the case in SR and still in GR:
\be\label{Ass1}
\mathrm{(A)\ Total\ energy\ tensor}\ \Couleur{\Mat{T} = \Mat{T}_\mathrm{charged\ medium} + \Mat{T}_\mathrm{field}}. 
\ee
(The total energy tensor \Couleur{$\Mat{T}$} is the source of the gravitational field --- more precisely, in SET, that source is the component \Couleur{$T^{00}$} in the preferred reference frame of the theory; see Ref.~\cite{A35} for details.) The additivity (\ref{Ass1}) leads to a form of Maxwell's second group of equations in SET. \cite{A54} But that form of Maxwell's second group in SET predicts charge production/destruction at untenable rates, therefore it has to be {\it discarded}. \cite{A56}\ The additivity assumption (\ref{Ass1}) is contingent and may be abandoned. \hypertarget{Intro-T_Inter}{This means introducing} an ``interaction" energy tensor \Couleur{$\Mat{T}_\mathrm{inter}$}, such that 
\be\label{Tinter}
\Couleur{\Mat{T} = \Mat{T}_\mathrm{charged\ medium} + \Mat{T}_\mathrm{field}\ \underline{+ \Mat{T}_\mathrm{inter}}\,}. 
\ee
One then has to constrain the form of \Couleur{$\Mat{T}_\mathrm{inter}$} and to derive equations for it.

\section{Form of the interaction tensor}\label{Intro-Inter}

In SR, the additivity (\ref{Ass1}) of the energy tensors does apply, thus \Couleur{$\Mat{T}_\mathrm{inter}={\bf 0}$}. In SET we may impose that \Couleur{$\Mat{T}_\mathrm{inter}$} should be Lorentz-invariant in the situation of SR, i.e. when the metric \Couleur{$\Mat{\gamma}$} is Minkowski's metric \Couleur{$\Mat{\gamma}^0$} (\Couleur{$\gamma^0 _{\mu \nu }=\eta _{\mu \nu }$} in Cartesian coordinates). This is true if and, one can prove, \, \cite{A58} {\it only if} we have: 
\be\label{T_inter_SR}
\Couleur{T_{\mathrm{inter}\ \mu \nu }= p\, \gamma^0 _{\mu \nu }}\qquad \mathrm{(situation\ of\ SR)},
\ee
with some scalar field \Couleur{$p$}. This is equivalent to:
\be\label{T_inter_SR_mixed}
\Couleur{T^\mu_{\mathrm{inter}\ \ \nu }= p\,\delta ^\mu _\nu} \qquad \mathrm{(situation\ of\ SR)}.
\ee
The definition
\be\label{T_inter_mixed}
\Couleur{T^\mu_{\mathrm{inter}\ \ \nu }:=  p\,\delta ^\mu _\nu}, \qquad \mathrm{or}\quad \Couleur{(T_\mathrm{inter})_{\mu  \nu }:=  p\,\gamma  _{\mu \nu}},
\ee
thus got in a Minkowski spacetime, is in fact generally-covariant. Hence, we adopt (\ref{T_inter_mixed}) for the general case. With a general metric \Couleur{$\Mat{\gamma}$}, the tensor (\ref{T_inter_mixed}) is still pointwise Lorentz-invariant --- in the sense that we have \Couleur{$(T_\mathrm{inter})_{\mu \nu }\,(X)= p(X)\, \eta _{\mu \nu }$} in any coordinates that are Cartesian at a given event \Couleur{$X$}, and this form remains invariant after any coordinate transformation that is Lorentz at \Couleur{$X$}, i.e., such that the matrix \Couleur{$\left(\frac{\partial x'^\mu}{\partial x^\nu}(X)\right)$} belongs to the Lorentz group.



\section{SET electrodynamics with the interaction tensor}

With the additivity assumption (\ref{Ass1}) of the energy tensors, i.e., \Couleur{$\Mat{T}_\mathrm{inter}={\bf 0}$}, the system of equations of electrodynamics of SET is closed, but violates charge conservation. With the interaction energy tensor (\ref{T_inter_mixed}) we have just one unknown more: the scalar field \Couleur{$p$}. So we need just one scalar equation more. It turns out to be consistent to add {\it charge conservation} as the new scalar equation. \cite{A57} Then the system of equations of electrodynamics of SET is again closed, and now it satisfies charge conservation.\\

Based on that closed system, equations were derived that {\it determine the field \Couleur{$p$} in a given general electromagnetic (EM) field \Couleur{$({\bf E},{\bf B})$} and in a given weak gravitational field with Newtonian potential \Couleur{$U$}}:  \cite{A57} the scalar field \Couleur{$p$} (or more exactly, its first approximation \Couleur{$p_1$}) obeys an advection equation:
\be\label{advec_p}
\Couleur{\partial _T\, p_1 + u^j \partial _j\,p_1 = S}.
\ee
That equation has given source \Couleur{$S$} and given characteristic curves, the latter being the integral curves \Couleur{$\mathcal{C}(T_0, {\bf x}_0)$} of the spatial vector field \Couleur{${\bf u}$} in Eq. (\ref{advec_p}). Here, ``given" means that the source field \Couleur{$S$}, as also the vector field \Couleur{${\bf u}$} and hence the characteristic curves \Couleur{$\mathcal{C}(T_0, {\bf x}_0)$}, do not depend on the unknown field \Couleur{$p_1$}. It follows that \Couleur{$p_1$} can be obtained by integrating the source field \Couleur{$S$} along those curves. \cite{A57}  \\

The ``medium" defined by the corresponding interaction energy tensor field \Couleur{$\Mat{T}_\mathrm{inter} = p \Mat{\gamma}$} can be counted as ``dark matter", for 
\bi
\item it is not localized inside (usual) matter: indeed, the equations for the field \Couleur{$p_1$} show that its source \Couleur{$S$} is, in general, non-zero as soon as there is a general EM field: \Couleur{${\bf E}\ne 0,\ {\bf B} \ne 0,\ {\bf E.B} \ne 0$}, and a variable gravitational field with \Couleur{$\partial _T U  \ne 0$}, where the time derivative \Couleur{$\partial _T U $} of the Newtonian potential is taken in the preferred frame; \cite{A57} \\

\item it is gravitationally active, since, from its definition (\ref{Tinter}), it contributes to the source of the gravitational field in SET, that is the component \Couleur{$T^{0 0}$} in the preferred frame;\\

\item it is ``exotic", i.e., it is not usual matter --- as shown by the form (\ref{T_inter_mixed}) of its energy tensor, which is very different from the possible energy tensors of any fluid, solid, or EM field. The fact that it is Lorentz-invariant means that no velocity can be defined for that medium. The energy tensor (\ref{T_inter_mixed}) depends only on one scalar field (\Couleur{$p$}), hence no equation of state is needed.
\ei 

The foregoing considerations are at the classical level, hence do not tell if the ``matter" with the energy tensor (\ref{T_inter_mixed}) is made of quantum particles.



\section{Maxwell model of the ISRF}

In order to check if the interaction energy \Couleur{$E_\mathrm{inter}$} might be distributed in the form of dark halos and contribute significantly to the dark matter distribution, we have to compute the field \Couleur{$p$} for a model of a galaxy. This needs that we have a model of the Interstellar Radiation Field in a galaxy (ISRF) that provides that field as a solution of the Maxwell equations. However, the existing models of the ISRF (e.g. Refs. \cite{Draine1978, Mathis-et-al1983, Chi-Wolfendale1991, Gordon-et-al2001, Robitaille2011, Popescu-et-al2017}) focus on the radiation transfer (mainly via absorption, reemission or scattering by dust particles). They follow the paths of light rays or photons. To the best of our knowledge, no previous model of the ISRF did consider the full EM field with its six interacting components subjected to the Maxwell equations. Therefore, we had to build a model entirely from scratch, which involved both theoretical and numerical difficulties. \cite{A61}

\subsection{Maxwell model of the ISRF: Main assumptions}

i) {\it Axial symmetry} is a relevant approximation for many galaxies, and is in fact often used in the existing models of the ISRF (see e.g. Refs. \cite{Popescu-et-al2017, KylafisBahcall1987, PorterStrong2005,Popescu-et-al2011}). We adopt cylindrical coordinates \Couleur{$(\rho ,\phi ,z)$} whose the \Couleur{$z$} axis is the symmetry axis. \\

The primary source of the ISRF is made of the stars or other bright astrophysical objects. We want to describe the ISRF as a smoothed-out field at the galactic scale, not the field in the stars or in their neighborhood. Therefore:\\

\noi ii) we consider the {\it source-free} Maxwell equations. \\

We proved the following result: \cite{A60}\\

\hypertarget{Theorem}{{\it Theorem.}} Any time-harmonic axisymmetric source-free Maxwell field is the sum of two simple fields of that same kind:
\bi
\item \hypertarget{field(1)}{{\bf 1}) one deriving from a vector potential} \Couleur{${\bf A}$} having just \Couleur{$A_z \ne 0$}, with \Couleur{$A_z$} a time-harmonic axisymmetric solution of the scalar wave equation; \\

\item \hypertarget{field(2)}{{\bf 2}) one deduced from a field of the form} ({\bf 1}) by EM duality, i.e.
\be\label{dual}
\Couleur{{\bf E}' = c{\bf B}, \quad {\bf B}' = -{\bf E}/c}.
\ee
\ei

\subsection{Maxwell model of the ISRF: Form of the model}\label{Form of Model}

We consider an EM field having a finite set of frequencies \Couleur{$(\omega _j)\ (j=1,...,N_\omega )$}. That EM field is thus the sum of $N_\omega $ time-harmonic EM fields. Using the \hyperlink{Theorem}{Theorem above}, each of them is generated by potentials \Couleur{$A_{j z},\,A'_{j z}$}. The scalar potential \Couleur{$A_{j z}$}, for the field of the form \hyperlink{field(1)}{(1) above} with frequency \Couleur{$\omega _j$}, can be a priori any time-harmonic axisymmetric solution of the scalar wave equation having frequency \Couleur{$\omega _j$}. \cite{A60} However, in the relevant ``totally propagating" case, such a solution can be written explicitly in terms of a spectrum function \Couleur{$S_j=S_j(k)$}\quad (\Couleur{$-K_j \le k \le K_j$},\quad \Couleur{$K_j:=\frac{\omega _j}{c}$}): \ \Couleur{$A_{j z}=\psi _{\omega _j\,S_j}$}, with \cite{ZR_et_al2008}
\be\label{psi_monochrom_j}
\Couleur{\psi _{\omega_j\ S_j} \,(t,\rho,z) := e^{-\iC \omega_j t} \int _{-K_j} ^{K_j}\ J_0\left(\rho \sqrt{K_j^2-k^2}\right )\ e^{\iC k \, z} \,S_j(k)\, \dd k},
\ee
where \Couleur{$J_0$} is the Bessel function of the first kind and of order 0. The ``dual" potential \Couleur{$A'_{j z}$}, for the field of the form \hyperlink{field(2)}{(2) above} with frequency \Couleur{$\omega _j$}, has just the same form (\ref{psi_monochrom_j}), with, in the general case, another spectrum function, say \Couleur{$S'_j$}.


\subsection{Maxwell model of the ISRF: Model of a galaxy}

We model an axisymmetric galaxy as a finite set \Couleur{$\{{\bf x}_i \} $} of point-like ``stars", the azimuthal distribution of which is uniform. That set of points is obtained by pseudo-random generation of their cylindrical coordinates \Couleur{$\rho ,\phi ,z$} with specific probability laws, ensuring that \cite{A61}
\bi
\item the distribution of \Couleur{$\rho $} and \Couleur{$z$} is approximately that valid for the star distribution in the galaxy considered (in the numerical application, we took our Galaxy); 

\vspace{1mm}
\item the set \Couleur{$\{{\bf x}_i \} $} is approximately invariant under azimuthal rotations of any angle \Couleur{$\phi $}.
\ei

\subsection{Maxwell model of the ISRF: Determining the potentials}

To determine the potentials \Couleur{$A_{j\,z}$} and \Couleur{$A'_{j\,z}$} (\Couleur{$j=1,..., N_\omega $}) that generate the model ISRF (Subsect. \ref{Form of Model}), we are fitting to the form (\ref{psi_monochrom_j}) a sum of spherical potentials emanating from the ``stars" at points \Couleur{${\bf x}_i$}, thus determining the unknown spectrum functions \Couleur{$S_j$} and  \Couleur{$S'_j$}. \cite{A61} For the purpose of this fitting, every point-like ``star" is indeed assumed to contribute spherical scalar waves \Couleur{$\psi _{{\bf x}_i\,\omega_j}$} having the same frequencies \Couleur{$\omega _j$} as has the model ISRF, and whose emission center is the spatial position \Couleur{${\bf x}_i$} of the star: 
\be\label{psi_spher_i} 
\Couleur{\psi _{{\bf x}_i\,\omega_j} \ (t,{\bf x}) := \psi _{\omega_j} \ (t,{\bf x}-{\bf x}_i) = \frac{e^{\iC (K_j \,r_i-\omega_j t)}} {K_j\, r_i}}.  
\ee
Here \Couleur{$r_i:=\abs{{\bf x}-{\bf x}_i}$}, \ \Couleur{$K_j:=\frac{\omega_j }{c}$}, and the function
\be\label{psi_spher}
\Couleur{\psi _{\omega_j} \ (t,{\bf x}) = \frac{e^{\iC (K_j\, r-\omega_j t)}} {K_j \,r}}, \qquad \Couleur{r:=\abs{{\bf x}}}
\ee
is (up to an amplitude factor) the unique time-harmonic solution of the scalar wave equation, with frequency \Couleur{$\omega _j$}, that has spherical symmetry around \Couleur{${\bf x} = {\bf 0}$} and that is an outgoing wave. Spherical symmetry is assumed in order to ensure that all of the directions starting from the star are equivalent, of course. Of course also, the uniqueness of the solution (\ref{psi_spher}) means the uniqueness of the solution translated from \Couleur{${\bf x} = {\bf 0}$} to \Couleur{${\bf x} = {\bf x}_i$}: the function $\Couleur{\psi _{{\bf x}_i\,\omega_j} }$  given by Eq. (\ref{psi_spher_i}). This implies that we cannot define different contributions of the ``star" at \Couleur{${\bf x}_i$} to the \Couleur{$A_{j\,z}$} potential and to the ``dual" potential \Couleur{$A'_{j\,z}$}, other than through multiplying $\Couleur{\psi _{{\bf x}_i\,\omega_j} }$ by two different amplitude factors --- for which there is no apparent reason. Therefore, we actually must assume that \Couleur{$A_{j\,z}=A'_{j\,z}$}, thus \Couleur{$S_j=S'_j$}, and our fitting problem writes
\be\label{Psi-simeq-Psi'-j-by-j-1}
\Couleur{\sum_{i=1} ^{i_\mathrm{max}} \psi _{{\bf x}_i\,\omega_j} \cong \psi _{\omega_j\ S_j}}\quad \mathrm{on}\ \Couleur{G}\qquad (\Couleur{j=1,..., N_\omega} ).
\ee
Here the symbol \Couleur{$\cong$} indicates that the equality is in the sense of the least-squares, the 
two sides being evaluated on some spatio-temporal grid \Couleur{$G$}. The unknown spectrum function \Couleur{$S_j$} is defined (approximately) by its values \Couleur{$S_{n j} := S_j(k_{n j})$} at a regular discretization \Couleur{$k_{n j} = -K_j + n\frac{2K_j}{N} \ (n=0,...,N)$} of the integration interval \Couleur{$[-K_j,+K_j]$} for \Couleur{$k$} in the integral (\ref{psi_monochrom_j}). \cite{A61} With this discretization, (\ref{Psi-simeq-Psi'-j-by-j-1}) becomes the explicit least-squares system
\be\label{Psi-simeq-Psi'-j-by-j-2}
\Couleur{\sum_{i=1} ^{i_\mathrm{max}} \psi _{{\bf x}_i\,\omega_j} \cong \sum _{n=0} ^N f_{n j} \,S_{n j}}\quad \mathrm{on}\ \Couleur{G} \Couleur{\qquad (j=1,...,N_\omega )},
\ee
with \Couleur{$f_{n j}(t,\rho ,z)=\exp(-\iC\omega _j t)\,g_{n j}(\rho ,z)$} a specific time-harmonic function. \cite{A63} The complex numbers \Couleur{$S_{n j}\ (n=0,...,N;\, j=1,...,N_\omega)$} are the solved-for parameters. Note that (\ref{Psi-simeq-Psi'-j-by-j-2}) defines \Couleur{$N_\omega $} fitting problems. Previously, a unique ``grouped fitting" was done: solving the least-squares system obtained by summing on the frequency index \Couleur{$j$} on both sides of (\ref{Psi-simeq-Psi'-j-by-j-2}). \cite{A61} The ``separate fitting" (\ref{Psi-simeq-Psi'-j-by-j-2}) is more precise --- and also less time-consuming, since actually the common harmonic time dependence can be removed from both sides of (\ref{Psi-simeq-Psi'-j-by-j-2}), thus eliminating the time variable, and hence considering only a spatial grid $G'$ instead of a spatio-temporal grid $G$. \cite{A63} The computer time is indeed an important point to be considered, because a precision better than quadruple must be implemented. \cite{A61}

\subsection{Application to the spatial variation of the spectral energy density in the Galaxy}

Because we consider an EM field with a finite frequency spectrum \Couleur{$(\omega _j)\ (j=1,...,N_\omega )$}, each among its six components has the following form:
\be\label{F(t)}
\Couleur{F^{(q)}(t,{\bf x}) = {\mathcal Re} \left ( \sum _{j=1} ^{N_\omega } C^{(q)}_j({\bf x}) e^{-\iC \omega _j t} \right )\qquad (q=1,...,6)}.
\ee
It follows that the time-averaged volumic energy density of the field is given by: \cite{A62}
\be\label{Udiscrete}
\Couleur{\overline{U}({\bf x}) :=\overline{\frac{\delta W}{\delta V}}({\bf x}) = \sum _{j=1} ^{N_\omega } u_j({\bf x}), \qquad u_j({\bf x}):= \frac{1}{4} \sum _{q=1} ^6 \alpha _q \abs{C^{(q)}_j({\bf x})}^2},
\ee
where \Couleur{$\alpha _q= \epsilon _0$} for an electric field component, whereas \Couleur{$\alpha _q= \epsilon _0 c^2$} for a magnetic field component (here \Couleur{$\epsilon _0 $} is the vacuum permittivity, with \Couleur{$\epsilon _0 = 1/(4\pi \times 9\times 10^9)$} in SI units). Thus, the spectral energy density (SED) has a discrete form. \\

\begin{figure}[tbp]
\centering
  \begin{minipage}[b]{0.45\textwidth}
    \includegraphics[width=\textwidth]{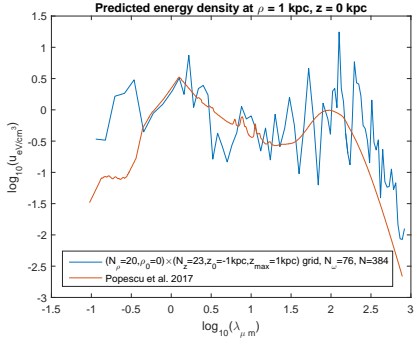}
  \end{minipage}
  \hfill
  \begin{minipage}[b]{0.45\textwidth}
    \includegraphics[width=\textwidth]{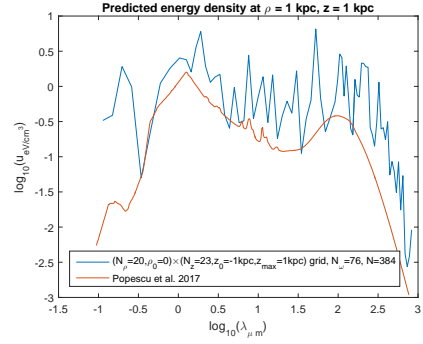}
  \end{minipage}
\caption{SEDs at ($\rho = 1$\,kpc, $z = 0$) and at ($\rho = 1$\,kpc, $z = 1\,$kpc).}
\label{SED(rho=1)}
\end{figure}

\begin{figure}[tbp]
\centering
  \begin{minipage}[b]{0.45\textwidth}
    \includegraphics[width=\textwidth]{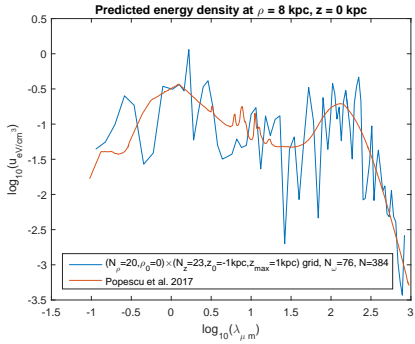}
  \end{minipage}
  \hfill
  \begin{minipage}[b]{0.45\textwidth}
    \includegraphics[width=\textwidth]{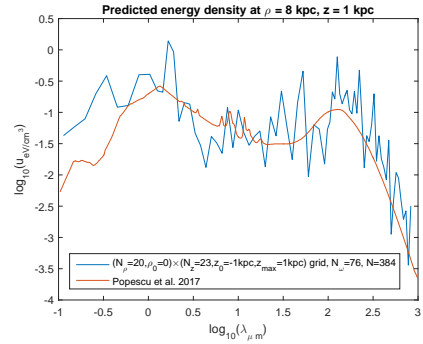}
  \end{minipage}
\caption{SEDs at ($\rho = 8$\,kpc, $z = 0$) and at ($\rho = 8$\,kpc, $z = 1\,$kpc).}
\label{SED(rho=8)}
\end{figure}

Specializing to the present axisymmetric model, we thus have \Couleur{$C^{(q)}_j= C^{(q)}_j(\rho ,z)$} and \Couleur{$u_j= u_j(\rho ,z)$}. The potentials \Couleur{$A_{j z} = A'_{j z} = \psi _{\omega _j\,S_j}$} are determined by the spectrum functions \Couleur{$S_j$} in Eq. (\ref{psi_monochrom_j}), which are given in the numerical model by the values \Couleur{$S_{n j} := S_j(k_{n j})$}, that are the output of the fitting. These potentials generate the EM field, hence the \Couleur{$C^{(q)}_j(\rho ,z)$} coefficients in Eq. (\ref{F(t)}) are expressed uniquely in terms of the \Couleur{$S_{n j}$} 's. \cite{A62} However, in the least-squares problem (\ref{Psi-simeq-Psi'-j-by-j-2}), the scalar radiations emitted by every point-like ``star" are taken to be exactly \Couleur{$\psi _{{\bf x}_i\,\omega_j}$}. Clearly, we may multiply the l.h.s. of (\ref{Psi-simeq-Psi'-j-by-j-2}) by some number \Couleur{$\xi _j>0$}, thus obtaining now new values \Couleur{$S'_{n j} =\xi _j S_{n j}\ (n=0,...,N)$} as the solution of (\ref{Psi-simeq-Psi'-j-by-j-2}). We determine the numbers \Couleur{$\xi _j>0$} so that the SED measured at our local position \Couleur{${\bf x}_\mathrm{loc}$} in the Galaxy coincides with the calculated values \Couleur{$u_j({\bf x}_\mathrm{loc})$}. This allows us then to make predictions: in particular, ones of the spatial variation of the SED in the Galaxy, which we may compare with the predictions of the existing models of the ISRF. Figures \ref{SED(rho=1)}--\ref{SED(rho=8)} show this comparison for the four positions in the Galaxy for which the predicted SED is shown in Ref. \cite{Popescu-et-al2017}. The predictions of the two models are quite reasonably close, although the SED predicted by the present model has rather marked oscillations as function of the wavelength. Note that the different wavelengths are fully uncoupled due to the ``separate fitting" defined by the \Couleur{$N_\omega $} least-square problems (\ref{Psi-simeq-Psi'-j-by-j-2}).\\

\begin{figure}[ht]
\centerline{\includegraphics[height=9cm]{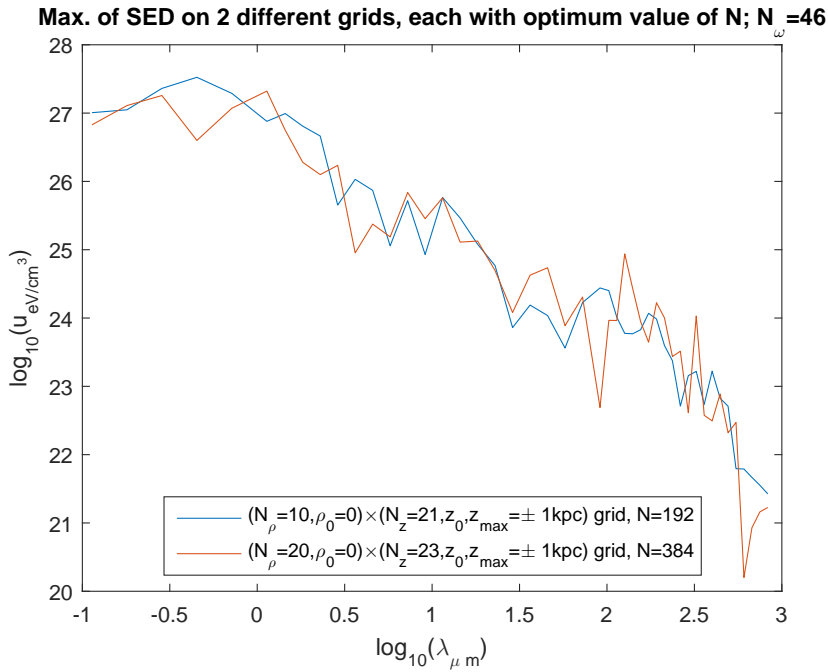}}
\caption{Maximum of energy density; comparison between two spatial grids.}
\label{uj_max_46-rough-vs-fine}
\end{figure}

A surprising prediction of this model is that for the values of the maximum of the energy density,
\be
\Couleur{u_{j \mathrm{max}} = \mathrm{Max}\{u_j(\rho _m, z_p); \ m=1,...,N_\rho ,\  p=1,...,N_z\}},
\ee
found for the different spatial grids (\Couleur{$N_\rho \times N_z$}) investigated, all having \Couleur{$\rho $} varying regularly from \Couleur{$\rho _0=0$} to \Couleur{$\rho_\mathrm{max}\simeq 10$}\,kpc and \Couleur{$z$} varying regularly from \Couleur{$z_0=0$} or \Couleur{$z_0=-z_\mathrm{max}$} to \Couleur{$z_\mathrm{max}\le 1\,$}kpc. Figure \ref{uj_max_46-rough-vs-fine} compares the curves 
\Couleur{$u_{j \mathrm{max}}=f(\lambda _j)$} found with two spatial grids. It is seen that the two curves are quite close to one another, and both show extremely high levels of \Couleur{$u_{j \mathrm{max}}$}, from $10^{27} \mathrm{eV/cm}^3$ to $10^{21} \mathrm{eV/cm}^3$. This is confirmed by a rather detailed investigation of the effects of the settings of the calculation (the spatial grid, and also the fineness of the frequency mesh: \Couleur{$N_\omega$}, and that of the discretization: \Couleur{$N$}). \cite{A63} The values of the maximum of \Couleur{$u_j$} are always found on the axis of the Galaxy (\Couleur{$\rho =0$}), moreover the level of \Couleur{$u_j$} decreases very rapidly when one departs from the axis. \cite{A63} This prediction of the model may be described as a kind of self-focusing effect of the ISRF in an axisymmetric galaxy. 

\section{Conclusion}

In the ``scalar ether theory" of gravity (SET), a consistent electrodynamics in a gravitational field needs the introduction of an additional energy tensor: \Couleur{$\Mat{T}_\mathrm{inter} $}, with \Couleur{$T^\mu_{\mathrm{inter}\ \ \nu }:=  p\,\delta ^\mu _\nu$}. Thus, this energy tensor was not designed to build missing mass. However, it turns out that the corresponding ``medium" could contribute to dark matter, for it is not localized inside matter, it is gravitationally active, and it is ``exotic". Moreover, the scalar field \Couleur{$p$}, that determines \Couleur{$\Mat{T}_\mathrm{inter} $}, can be in principle calculated from the data of the EM field and the gravitational field, through explicit equations. This however demands to be able to model the EM field in a galaxy, which is essentially in the form of the interstellar radiation field (ISRF). \\

Therefore, we built a Maxwell model of the ISRF. This was motivated by the foregoing, but it is also interesting independently of that, as the ISRF is a very important physical characteristic of a galaxy and interacts strongly with the cosmic rays. In any case, this model in itself is totally independent of the theory of gravitation and the assumption about the interaction tensor. It is based on an explicit representation of any time-harmonic axisymmetric source-free Maxwell field through a pair of scalar potentials, and on determining these potentials by fitting contributions emanating from a set of point-like ``stars" schematizing a galaxy. The predictions of the model for the variation of the spectral energy distribution in the Galaxy are currently being checked. They are relatively close to the predictions of a recent radiation transfer model --- except for the fact that the Maxwell model of the ISRF predicts extremely high values of the energy density on the axis of the Galaxy, that however decrease very rapidly when departing from that axis. We hope to be able in a future work to apply the model to calculate the interaction energy and to check if its distribution resembles a dark halo.



\end{document}